\documentclass[12pt]{article}
\usepackage{bbm}
\newcommand{\be}{\begin{equation}}
\newcommand{\ee}{\end{equation}}
\newcommand{\ba}{\begin{eqnarray}}
\newcommand{\ea}{\end{eqnarray}}
\newcommand{\no}{\nonumber\\}
\begin{document}
\title{\normalsize \hfill UWThPh-2003-6 \\[1cm] \LARGE
A discrete symmetry group \\
for maximal atmospheric neutrino mixing}
\author{Walter Grimus\thanks{E-mail: walter.grimus@univie.ac.at} \\
\setcounter{footnote}{3}
\small Institut f\"ur Theoretische Physik, Universit\"at Wien \\
\small Boltzmanngasse 5, A--1090 Wien, Austria \\*[3.6mm]
Lu\'{\i}s Lavoura\thanks{E-mail: balio@cfif.ist.utl.pt} \\
\small Universidade T\'ecnica de Lisboa \\
\small Centro de F\'\i sica das Interac\c c\~oes Fundamentais \\
\small Instituto Superior T\'ecnico, P--1049-001 Lisboa, Portugal \\*[4.6mm] }

\date{1 August 2003}

\maketitle

\begin{abstract}
We propose a discrete non-abelian symmetry group
which enforces maximal atmospheric neutrino mixing,
while $\theta_{13} = 0$
and the solar mixing angle  $\theta_{12}$ remains undetermined;
without finetuning,
$\theta_{12}$ will be large but non-maximal.
Our extension of the Standard Model
has three right-handed neutrino singlets $\nu_R$
and implements the seesaw mechanism.
Furthermore,
we have an enlarged scalar sector with three Higgs doublets
and two scalar gauge singlets;
the latter have masses and vacuum expectation values
of the order of the seesaw scale.
Lepton mixing stems exclusively from the $\nu_R$ Majorana mass matrix, 
where non-diagonal elements are generated
by the vacuum expectation values of the scalar singlets.
The model predicts a neutrino mass spectrum with $m_3 > m_2 > m_1$,
and the effective Majorana mass of neutrinoless $\beta\beta$ decay
is equal to $m_1 m_2 / m_3$.
\end{abstract}

\newpage

\section{Introduction}

The KamLAND experiment \cite{kamland}
has confirmed that the solar neutrino deficit---for a recent review see
Ref.~\cite{goswami}---is explained by neutrino oscillations \cite{pontecorvo},
with matter effects \cite{MSW} playing a decisive role.
Whereas the solution of the solar neutrino problem,
the large mixing angle MSW solution,
displays a large but non-maximal mixing angle $\theta_{12}$ \cite{goswami},
the atmospheric neutrino problem with mixing angle $\theta_{23}$
requires $\sin^2 2\theta_{23} > 0.92$ at 90\% CL,
with a best-fit value $\sin^2 2\theta_{23} = 1$ \cite{SK}.

There are many models and textures in the literature
which attempt to explain large (not necessarily maximal) 
atmospheric neutrino mixing---for reviews see Ref.~\cite{reviews}.
However,
the closer the experimental lower bound 
on $\sin^2 2\theta_{23}$ comes to 1,
the more urgent it becomes
to find a rationale for maximal atmospheric neutrino mixing. 
Arguments have been given \cite{nussinov,wetterich,king}
that in this case a non-abelian symmetry
is required---for models along this line see,
for instance,
Refs.~\cite{GL01,ohlsson,ma,kitabayashi}.

In this letter we propose a model
which employs a discrete non-abelian symmetry group $G$ with eight elements.
The essential ingredients of this model
are the seesaw mechanism \cite{seesaw}
and a pair of scalar gauge singlets $\chi_{1,2}$
in a two-dimensional irreducible representation of $G$.
Neutrino mixing stems solely from the Majorana mass matrix $M_R$
of the right-handed neutrino singlets $\nu_R$.
The vacuum expectation values (VEVs) of $\chi_{1,2}$
generate off-diagonal elements of $M_R$,
and atmospheric mixing turns out maximal.
It is essential for our model that the VEVs of the scalar singlets
be at the (large) seesaw scale.
The present model has some similarities with the model of Ref.~\cite{GL01},
where part of the symmetry group is given
by the non-abelian group $O(2)$ \cite{GLsu5}. 

This paper is organized as follows.
In Section~\ref{particle} we introduce the multiplets of the model
and in Section~\ref{group} we discuss the horizontal symmetry group $G$.
The Yukawa sector,
the VEVs of the scalar singlets,
and the neutrino masses and mixings are presented in Sections~\ref{yukawa},
\ref{vev},
and \ref{massmatrix},
respectively;
in Section~\ref{mutau} we digress to explain
how $m_\mu \ll m_\tau$ may be achieved in a technically natural way
in the context of our model.
After discussing some of the predictions of the model in Section~\ref{phen},
we come to the conclusions in Section~\ref{concl}.

\section{Particle content and symmetries}
\label{particle}

Our model has three lepton families,
including three right-handed neutrinos.
We generically denote $e$,
$\mu$,
and $\tau$ by $\alpha$.
Thus,
we have three left-handed lepton doublets $D_\alpha$,
three right-handed charged-lepton singlets $\alpha_R$,
and three right-handed neutrino singlets $\nu_{\alpha R}$.

In the scalar sector,
we employ three Higgs doublets $\phi_1$,
$\phi_2$,
and $\phi_3$,
together with two neutral singlets $\chi_1$ and $\chi_2$.
For the sake of simplicity we shall assume the singlets to be real.

We introduce three symmetries of the $\mathbbm{Z}_2$ type:
\be\label{symm}
\begin{array}{rl}
\mathbbm{Z}_2^{(\tau)}: &
D_\tau,\
\tau_R,\
\nu_{\tau R},\
\chi_2\, \
\mathrm{change\ sign;}
\\
\mathbbm{Z}_2^{(\mathrm{tr})}: &
D_\mu \leftrightarrow D_\tau,\
\mu_R \leftrightarrow \tau_R,\
\nu_{\mu R} \leftrightarrow \nu_{\tau R},\
\chi_1 \leftrightarrow \chi_2,\ 
\phi_3 \to - \phi_3; 
\\
\mathbbm{Z}_2^{(\mathrm{aux})}: &
\nu_{eR},\
\nu_{\mu R},\
\nu_{\tau R},\
\phi_1,\ 
e_R\, \
\mathrm{change\ sign.}
\end{array}
\ee
$\mathbbm{Z}_2^{(\tau)}$ flips the signs of all multiplets
of the $\tau$ family
and $\mathbbm{Z}_2^{(\mathrm{tr})}$
makes a transposition of the multiplets of the $\mu$ and $\tau$ families. 
The auxiliary symmetry $\mathbbm{Z}_2^{(\mathrm{aux})}$ 
serves the purpose of allowing for $m_\mu \neq m_\tau$
while preserving the appropriate form
of the neutrino mass matrix $\mathcal{M}_\nu$,
as we shall see soon.
The auxiliary $\mathbbm{Z}_2$
is equivalent to the one employed in the model of Ref.~\cite{GL01},
whose charged-lepton mass generation is identical with the present one.

\section{The group}
\label{group}

The symmetries $\mathbbm{Z}_2^{(\tau)}$ and $\mathbbm{Z}_2^{(\mathrm{tr})}$
do not commute,
and together they generate a non-abelian group $G$ with eight elements:
\be
{\rm set} \left( G \right) = \left\{ e, g, h, g h, h g, g h g, h g h,
g h g h = h g h g \right\},
\ee
where $e$ is the identity element,
and $g$ and $h$ satisfy
\be
g^2 = h^2 = \left( g h \right)^4 = e.
\ee
The group $G$ has five inequivalent irreducible representations:
\be
\begin{array}{rl}
\underline{2}: &
g \to \left( \begin{array}{cc} 1 & 0 \\ 0 & - 1
\end{array} \right),\
h \to \left( \begin{array}{cc} 0 & 1 \\ 1 & 0
\end{array} \right);
\\[1mm]
\underline{1}_{\sigma \sigma^\prime}: &
g \to \sigma 1,\ h \to \sigma^\prime 1,
\ \mathrm{with} \ \sigma, \sigma^\prime = \pm.
\end{array}
\ee
The $\underline{2}$ is the only faithful irreducible representation.
It is obvious that $\left( D_\mu, D_\tau \right)$,
$\left( \mu_R, \tau_R \right)$,
$\left( \nu_{\mu R}, \nu_{\tau R} \right)$,
and $\left( \chi_1, \chi_2 \right)$ transform as $\underline{2}$,
while $\phi_3$ is a $\underline{1}_{+-}$.

The irreducible representations of the group $G$
have a single non-trivial tensor product:
\be
\underline{2} \otimes \underline{2} =
\underline{1}_{++} \oplus
\underline{1}_{+-} \oplus
\underline{1}_{-+} \oplus
\underline{1}_{--}.
\ee

From the $\underline{2}$ we read off
that $G$ can be thought of as a group of symmetry operations in the plane
comprising rotations with multiples of $90^\circ$
and reflections at the $x$ and $y$ axes
and at the axes obtained by rotating the coordinate axes
by an angle of $45^\circ$.
This group is isomorphic to the crystallographic point group $D_4$
\cite{cornwell}.

\section{Yukawa couplings}
\label{yukawa}

The Yukawa Lagrangian determined by the multiplets and symmetries
defined in Section~\ref{particle} is given by
\ba
\mathcal{L}_\mathrm{Y} &=&
- \left[ y_1 \bar D_e \nu_{eR}
+ y_2 \left( \bar D_\mu \nu_{\mu R} + \bar D_\tau \nu_{\tau R} \right) \right]
\tilde \phi_1
\no & &
- y_3 \bar D_e e_R \phi_1
- y_4 \left( \bar D_\mu \mu_R + \bar D_\tau \tau_R \right) \phi_2
- y_5 \left( \bar D_\mu \mu_R - \bar D_\tau \tau_R \right) \phi_3
\no & &
+ \frac{y_\chi}{2}\, \nu_{eR}^T C^{-1} 
\left( \nu_{\mu R} \chi_1 + \nu_{\tau R} \chi_2 \right)
+ \mbox{H.c.},
\label{Y}
\ea
where $\tilde \phi_1 = i \tau_2 \phi_1^\ast$.
With VEVs $\left\langle 0 \left| \phi_j^0 \right| 0 \right\rangle
= v_j \left/ \sqrt{2} \right.$
($j = 1, 2, 3$),
one obtains $\sqrt{2}\, m_e = \left| y_3 v_1 \right|$ and
\be
m_\mu  = \frac{1}{\sqrt{2}} \left| y_4 v_2 + y_5 v_3 \right|,
\quad
m_\tau = \frac{1}{\sqrt{2}} \left| y_4 v_2 - y_5 v_3 \right|.
\label{masses}
\ee
Also,
the neutrino Dirac mass matrix is 
\begin{equation}\label{MD}
M_D = {\rm diag} \left( a, b, b \right), \;
\mathrm{with} \; a = y_1^\ast v_1 \left/ \sqrt{2} \right.
\; \mathrm{and} \;
b = y_2^\ast v_1 \left/ \sqrt{2} \right. .
\end{equation}
The smallness of the electron mass $m_e$
is correlated with the smallness of the neutrino masses,
since $m_e$ is proportional to $\left| v_1 \right|$
while the neutrino masses are,
through the seesaw mechanism,
proportional to $\left| v_1 \right|^2$.

Note that 
$v = \sqrt{|v_1|^2 + |v_2|^2 + |v_3|^2} \simeq 246\, \mathrm{GeV}$
represents the electroweak scale.

There is also a Majorana mass term
\be\label{majorana}
\mathcal{L}_\mathrm{M} = \frac{1}{2} \left[
M^\ast \nu_{eR}^T C^{-1} \nu_{eR} +
{M^\prime}^\ast \left( \nu_{\mu R}^T C^{-1} \nu_{\mu R}
+ \nu_{\tau R}^T C^{-1} \nu_{\tau R} \right) \right] + \mbox{H.c.}
\ee

Since in our model the charged-lepton mass matrix and $M_D$ are both diagonal,
the lepton mixing matrix has to result exclusively from $M_R$.
Indeed,
non-zero VEVs of $\chi_{1,2}$ will render $M_R$ non-diagonal.
Furthermore,
examining the Lagrangians of Eqs.~(\ref{Y}) and~(\ref{majorana}),
we find accidental $\mathbbm{Z}_2$ symmetries  
$\mathbbm{Z}_2^{(e)}$ and $\mathbbm{Z}_2^{(\mu)}$,
defined in analogy to $\mathbbm{Z}_2^{(\tau)}$,
cf.\ Eq.~(\ref{symm}):
\begin{equation}\label{acc}
\begin{array}{rl}
\mathbbm{Z}_2^{(e)}: &
D_e,\
e_R,\
\nu_{e R},\ 
\chi_1,\
\chi_2\, \
\mathrm{change\ sign;}
\\
\mathbbm{Z}_2^{(\mu)}: &
D_\mu,\
\mu_R,\
\nu_{\mu R},\
\chi_1\, \
\mathrm{change\ sign.}
\end{array}
\end{equation}
The properties mentioned in this paragraph are reminiscent
of the maximal-atmospheric-mixing model of Ref.~\cite{GL01}. 
The $\mathbbm{Z}_2^{(\alpha)}$ symmetries of the present model
correspond to the $U(1)_{L_\alpha}$ lepton-number symmetries
of Ref.~\cite{GL01};
in the neutrino sector,
the former symmetries are spontaneously broken
by the VEVs of $\chi_{1,2}$ leading to non-diagonal contributions to $M_R$,
as we shall see in the next sections,
whereas the lepton-number symmetries of Ref.~\cite{GL01}
are softly broken by the matrix elements of $M_R$. 
We further remark that,
through a change in the basis
used for the representation $\underline{2}$ of the group $G$,
one can construct a model completely equivalent to the present one,
where atmospheric mixing originates in the charged-lepton sector
and solar mixing originates in the neutrino sector.

\section{The VEVs of the scalar gauge singlets} \label{vev}

We write these VEVs as
\be
\left\langle 0 \left| \chi_1 \right| 0 \right\rangle
= W \cos{\gamma} \; \; \mathrm{and} \
\left\langle 0 \left| \chi_2 \right| 0 \right\rangle = W \sin{\gamma},
\label{param1}
\ee
with $W > 0$.
In the scalar potential $V$ the $\chi$-dependent terms compatible
with the symmetries of Eq.~(\ref{symm}) are
\ba
V &=& \cdots - \mu \left( \chi_1^2 + \chi_2^2 \right) 
+ \left( \chi_1^2 + \chi_2^2 \right) \sum_{j=1}^3 
\rho_j\, \phi_j^\dagger \phi_j
+ \lambda \left( \chi_1^2 + \chi_2^2 \right)^2
\no & &
+ \left( \chi_1^2- \chi_2^2 \right)
\left( \eta \phi_2^\dagger \phi_3 + \eta^\ast \phi_3^\dagger \phi_2 \right)
+ \lambda^\prime \left( \chi_1^2- \chi_2^2 \right)^2,
\label{pot}
\ea
where $\mu$, 
$\rho_j$ ($j = 1,2,3$),
$\eta$,
$\lambda$,
and $\lambda^\prime$ are constants.
This potential is similar to the ones used in Ref.~\cite{babu}.
Note that a term $\left( \chi_1 \chi_2 \right)^2$
is implicitly contained in $V$.
Thus,
$\gamma$ is determined by the minimization of
\be
f \left( \gamma \right) = \lambda' W^4 \cos^2{2 \gamma} 
+ \mathrm{Re} \left( \eta v_2^\ast v_3 \right) W^2 \cos{2 \gamma}.
\ee
Provided $\lambda^\prime > 0$
and the coefficient of $\cos{2\gamma}$
is smaller than the coefficient of $\cos^2{2\gamma}$ by,
at least,
a factor of 2,
the minimum of $f$ is at
\be
\cos{2 \gamma} =
- \frac{\mathrm{Re} \left( \eta v_2^\ast v_3 \right)}
{2 \lambda^\prime W^2}\, .
\ee
According to the seesaw mechanism \cite{seesaw},
we assume that $W \sim |M|, \, |M^\prime| \gg v$.
Then, $\left| \cos{2 \gamma} \right|$ is very small.
From now on we shall make the simplifying assumption that
\begin{equation}\label{VEVchi}
\left\langle 0 \left| \chi_1 \right| 0 \right\rangle =
\left\langle 0 \left| \chi_2 \right| 0 \right\rangle =
\frac{W}{\sqrt{2}}\, .
\end{equation}
Corrections to this relation are of order $v^2 \left/ W^2 \right.$,
i.e.\ they are suppressed
by the square of the ratio of electroweak over seesaw scale.

The potential in Eq.~(\ref{pot}) is invariant
not only under the $\mathbbm{Z}_2$ symmetries of Eq.~(\ref{symm}),
but also under the accidental symmetries of Eq.~(\ref{acc}).

\section{A possible mechanism for $m_\mu/m_\tau \ll 1$}
\label{mutau}

The muon and tau masses are given by Eq.~(\ref{masses}).
We find nothing wrong with the finetuning needed
for obtaining $m_\mu \ll m_\tau$,
since an analogous finetuning is also needed in other models,
in particular in the Standard Model.
Yet,
one could argue that in this case
we do not simply have to choose one parameter to be small,
rather we have to choose two products of unrelated quantities
such that in $m_\mu$ those two products nearly cancel.

In order to give a natural reason why $m_\mu \ll m_\tau$,
we may add to our model the following additional symmetry:
\be
\label{T}
T: \quad \mu_R \to -\mu_R,
\quad 
\phi_2 \leftrightarrow  \phi_3.
\ee
This symmetry yields
\be
\label{yy}
y_4 = - y_5.
\ee
In this way the finetuning is confined to the VEVs:
\be
\frac{m_\mu}{m_\tau} = \left| \frac{v_2 - v_3}{v_2 + v_3} \right|.
\ee

One may show that a potential $V$ which is invariant under $T$,
in addition to the symmetries of Eq.~(\ref{symm}),
will in general allow for vacua which do not break $T$ spontaneously,
i.e.\ in which the VEVs of $\phi_2^0$ and $\phi_3^0$ are identical.
Indeed,
if we write
\be
\frac{v_1}{\sqrt{2}} = u_1\, ,
\;
\frac{v_2}{\sqrt{2}} = u e^{i\alpha} \cos{\sigma}\, ,
\;
\frac{v_3}{\sqrt{2}} = u e^{i\beta} \sin{\sigma}\, ,
\ee
with $u_1$ and $u$ real and non-negative ($u_1^2 + u^2 = v^2/2$),
and $\sigma$ belonging to the first quadrant without loss of generality,
we find that there is a range of the constants in the scalar potential
for which $\alpha = \beta$ and $\sigma = \pi / 4$,
i.e.\ $v_2 = v_3$.
Thus,
it is possible to obtain $m_\mu = 0$ by imposing the symmetry $T$
and by not allowing it to be spontaneously broken.

Now,
in order to obtain $m_\mu \neq 0$ but small compared to $m_\tau$,
it is enough to introduce a soft breaking of $T$ in the scalar potential.
It is easy to see that the only possible term breaking $T$ softly---but
keeping intact the other symmetries of the theory,
in particular the ones in Eq.~(\ref{symm})---is
\be
\mu_\mathrm{soft}
\left( \phi_2^\dagger \phi_2 - \phi_3^\dagger \phi_3 \right),
\ee
in which $\mu_\mathrm{soft}$ is a real coupling constant
with dimensions of squared mass,
which may \emph{naturally} be assumed to be much smaller than $v^2$.
It is then easy to show that in the vacuum one gets
$\cos{2 \sigma} \propto
\left. \mu_\mathrm{soft} \right/ u^2 \ll 1$,
i.e.\ $\left| \left| v_2 \right|^2 - \left| v_3 \right|^2 \right|
\ll \left| v_2 \right|^2 + \left| v_3 \right|^2$,
hence $0 \neq m_\mu \ll m_\tau$.

Thus,
in the context of our model,
and contrary to what happens in many other models,
in particular in the Standard Model,
it is possible to introduce a natural mechanism,
operating through a softly broken discrete symmetry,
for $m_\mu \ll m_\tau$.

\section{The neutrino mass matrix}
\label{massmatrix}

With Eqs.~(\ref{Y}),
(\ref{majorana}),
and (\ref{VEVchi}) it is straightforward to write down
the Majorana mass matrix of the right-handed neutrino singlets:
\begin{equation}\label{MR}
M_R = \left(
\begin{array}{ccc}
M & M_\chi & M_\chi \\
M_\chi & M' & 0 \\
M_\chi & 0 & M'
\end{array}
\right), \; \mathrm{with} \ 
M_\chi = \frac{y_\chi^\ast W}{\sqrt{2}}\, .
\end{equation}
Using the seesaw formula \cite{seesaw}
\begin{equation}
\mathcal{M}_\nu = - M_D^T M_R^{-1} M_D
\end{equation}
for the Majorana mass matrix of the light neutrinos,
it follows readily from the $\mathbbm{Z}_2^{(\mathrm{tr})}$ invariance
of $M_D$ and $M_R$---see Eqs.~(\ref{MD}) and (\ref{MR}), respectively---that
$\mathcal{M}_\nu$ has the structure \cite{GL01}
\begin{equation}\label{Mnu}
\mathcal{M}_\nu =
\left( \begin{array}{ccc}
x & y & y \\ y & z & w \\ y & w & z
\end{array} \right).
\end{equation}
Maximal atmospheric neutrino mixing and $\theta_{13} = 0$
immediately follow from this structure of $\mathcal{M}_\nu$---see,
for instance,
Ref.~\cite{lam}---since that matrix has an eigenvector $(0,1,-1)^T$. 
In our model,
the form of $\mathcal{M}_\nu$ is a consequence
of the non-abelian symmetry group $G$. 
The solar mixing angle is not fixed by the mass matrix of Eq.~(\ref{Mnu}),
but without finetuning it will be large and non-maximal \cite{GL01}. 

More specifically, 
the symmetric matrix of Eq.~(\ref{Mnu})
is diagonalized by a unitary matrix $V$ as 
\be
V^T {\cal M}_\nu V = {\rm diag} \left( m_1, m_2, m_3 \right),
\label{urtsk}
\ee
where the light-neutrino masses $m_1$,
$m_2$,
and $m_3$ are real and non-negative by definition.
The matrix $V$ is decomposed as
\be\label{V}
V = e^{i \hat \alpha} U
e^{i \hat \beta}, \; \mathrm{with} \
U =
\left( \begin{array}{ccc}
\cos \theta & \sin \theta & 0 \\
-\sin \theta/\sqrt{2} & \cos \theta/\sqrt{2} & \hphantom{-}1/\sqrt{2} \\
-\sin \theta/\sqrt{2} & \cos \theta/\sqrt{2} & -1/\sqrt{2}
\end{array} \right),
\ee
where $\theta \equiv \theta_{12}$ is the solar mixing angle and $\hat \alpha$,
$\hat \beta$ are diagonal phase matrices.
The above-mentioned eigenvector of $\mathcal{M}_\nu$
appears as the third column of $U$.
We assume,
without loss of generality,
that $\theta$ belongs to the first quadrant while $m_2 > m_1$.
The solar mass-squared difference is $\Delta m^2_\odot \equiv
\Delta m^2_{21} = m_2^2 - m_1^2$. 
The physical Majorana phases are $2 \left( \beta_2 - \beta_1 \right)$
and $2 \left( \beta_3 - \beta_2 \right)$;
the other phases in $V$,
notably the phases in the matrix $\hat \alpha$,
are unphysical.

\section{Phenomenological consequences of the model}
\label{phen}

Our model does not allow for a fully general matrix $M_R$,
since it requires $\left( M_R \right)_{\mu\tau} = 0$,
cf.\ Eq.~(\ref{MR}).
A short calculation shows that this condition is rewritten as
\begin{equation}\label{cond}
\sum_{j = 1}^3 U_{\mu j} U_{\tau j}\, m_j^{-1}\, e^{2i\beta_j} = 
\frac{1}{2} \left( \frac{\sin^2 \theta}{m_1}\, e^{2i\beta_1} +
\frac{\cos^2 \theta}{m_2}\, e^{2i\beta_2} -
\frac{1}{m_3}\, e^{2i\beta_3} \right) = 0.
\end{equation}
This relation allows one to express
\begin{equation} 
\left| \left\langle m \right\rangle \right|
= \left| \left( \mathcal{M}_\nu \right)_{ee} \right| = 
\left| m_1 \cos^2 \theta\, e^{2 i \beta_1} + 
m_2 \sin^2 \theta\, e^{2 i \beta_2} \right|, 
\end{equation}
the effective mass probed in neutrinoless
$\beta\beta$ decay,
in terms of the neutrino masses:
\begin{equation}\label{123}
\left| \left\langle m \right\rangle \right|
= \frac{m_1 m_2}{m_3} \,.
\end{equation}
Equation~(\ref{cond}) also states that,
in the complex plane,
it is possible to draw a triangle with sides of length $\sin^2 \theta / m_1$,
$\cos^2 \theta / m_2$,
and $1 / m_3$.
Therefore,
the sum of any two of these numbers must be larger than,
or equal to,
the third one.
One such inequality is
\begin{equation}\label{ineq1}
\frac{\sin^2 \theta}{m_1} + \frac{\cos^2 \theta}{m_2} \geq
\frac{1}{m_3} \,.
\end{equation}
Since we have chosen the convention $m_2 > m_1$,
it follows from Eq.~(\ref{ineq1}) that $m_3 \geq m_1$.
Thus,
our model requires a neutrino mass spectrum with $m_1 < m_2 < m_3$.
The remaining two triangle inequalities are summarized as
\begin{equation}\label{ineq2}
\frac{1}{m_3} \geq \left|
\frac{\sin^2 \theta}{m_1} - \frac{\cos^2 \theta}{m_2} \right|.
\end{equation}
We can rewrite Eq.~(\ref{ineq2}) in the form 
\begin{equation}\label{ineq3}
\frac{m_1}{\sqrt{m_1^2 +
\Delta m^2_{\rm atm}}}
\geq \left| \sin^2 \theta - 
\cos^2 \theta \frac{m_1}{\sqrt{m_1^2 +
\Delta m^2_\odot}}
\right|,
\end{equation}
where we have identified $\Delta m^2_{31} = m_3^2 - m_1^2$
with the atmospheric mass-squared difference $\Delta m^2_\mathrm{atm}$. 

Since $\theta$ is large and $\Delta m^2_{\rm atm} \gg \Delta m^2_\odot$,
it is clear that Eq.~(\ref{ineq3}) always holds
for $m_1 \gg \sqrt{\Delta m^2_{\rm atm}}$.
Thus,
our model allows for a quasi-degenerate neutrino mass spectrum.
As a matter of fact,
Eq.~(\ref{ineq3}) is satisfied
either when $m_1$ is larger than a certain value of order
$\sqrt{\Delta m^2_{\rm atm}}$,
or when $m_1$ is in the vicinity of a certain value of order
$\sqrt{\Delta m^2_\odot}$
for which the right-hand side of that inequality vanishes.
For the sake of concreteness,
let us take the best-fit values
$\Delta m^2_\mathrm{atm} = 2.5 \times 10^{-3}\, \mathrm{eV}^2$ \cite{SK}
and 
$\Delta m^2_\mathrm{\odot} = 7.17 \times 10^{-5}\, \mathrm{eV}^2$,
$\tan^2 \theta = 0.44$ \cite{goswami}.
Then,
we obtain from Eq.~(\ref{ineq3}) that either
\be
3.16 \times 10^{-3}\, \mathrm{eV}
\le m_1 \le 7.09 \times 10^{-3}\, \mathrm{eV}
\label{iutyp}
\ee
or
\be\label{r2}
m_1 \ge 1.64 \times 10^{-2}\, \mathrm{eV}.
\ee
The value of $\left| \left\langle m \right\rangle \right|$
increases monotonically with increasing $m_1$.
For the lowest $m_1$ in the range of Eq.~(\ref{iutyp})
one obtains $\left| \left\langle m \right\rangle \right|
= 5.7 \times 10^{-4}\, \mathrm{eV}$.
This value is too low to be within the range of sensitivity
of the forthcoming experiments on neutrinoless $\beta \beta$
decay---for a review see Ref.~\cite{vogel}.
However,
for the range of $m_1$ in Eq.~(\ref{r2}),
we have the lower bound $\left| \left\langle m \right\rangle \right|
\ge 5.8 \times 10^{-3}\, \mathrm{eV}$,
which is not far from future experimental sensitivity.

In our model all the Yukawa-coupling matrices are diagonal. 
In the low-energy sector,
where the effects of $\chi_{1,2}$ are given solely by their VEVs
and their respective contributions to $M_R$, 
the only source of lepton-flavor violation is neutrino mixing. 
Denoting the scale of the right-handed neutrino masses, 
i.e.\ the seesaw scale,
by $m_R$,
it has been shown in Ref.~\cite{GL02} that in such a theory,
in the limit $m_R \to \infty$ and in the case of two or more Higgs doublets,
the lepton-flavor-changing neutral-scalar vertices do not vanish,
i.e.,
there is no decoupling in the Yukawa sector in that limit;
the non-vanishing one-loop corrections
are generated by charged-scalar exchange
and by the mixing of the heavy neutrinos. 
On the other hand,
all the lepton-flavor-changing processes
which have either photon or $Z$ exchange,
or proceed via a box diagram,
are completely negligible,
since their amplitudes are suppressed by $1/m_R^2$,
and thus decouple in the above limit.
Processes which are mediated by neutral-scalar exchange
and remain unsuppressed by inverse powers of $m_R$
are $\tau^\pm \to \mu^\pm \mu^+ \mu^-$, 
$\tau^\pm \to \mu^\pm e^+ e^-$, 
$\tau^\pm \to e^\pm e^+ e^-$,
and $\mu^\pm \to e^\pm e^+ e^-$.
Their branching ratios have been estimated \cite{GL02}
to be of order $Y^8 / \left( 16 \pi^2 G_F M_0^2 \right)^2$, 
where $Y$ is a typical Yukawa coupling,
$G_F$ is the Fermi constant,
and $M_0$ is a typical mass
of the five ``light'' physical neutral scalars in our model.
Using the reasonable values $16 \pi^2 G_F M_0^2 \sim 10$ and $Y \sim 10^{-2}$,
such branching ratios are of order $10^{-18}$ \cite{GL02}. 
However,
due to the proportionality to $Y^8$, 
a slight increase in $Y$
could easily bring the branching ratio for $\mu^\pm \to e^\pm e^+ e^-$
close to the sensitivity range of future experiments.

\section{Conclusions}
\label{concl}

In this letter we have discussed an extension of the Standard Model
in the lepton sector
which employs a horizontal non-abelian discrete symmetry group $G$.
With respect to the Standard Model,
the new multiplets are three right-handed neutrino singlets,
which allow the use of the seesaw mechanism,
two additional Higgs doublets,
and two real scalar gauge singlets $\chi_{1,2}$.
The $G$-multiplets are such that all the Yukawa-coupling matrices
of the Higgs doublets are diagonal.
Before $\chi_{1,2}$ acquire VEVs,
the mass matrix $M_R$ of the right-handed neutrino singlets
is diagonal as well.
Thus,
non-trivial lepton mixing
originates in $\left\langle \chi_{1,2} \right\rangle_0 \neq 0$.
Maximal atmospheric mixing
and the vanishing of the mixing angle $\theta_{13}$
are linked to $\left\langle \chi_1 \right\rangle_0
= \left\langle \chi_2 \right\rangle_0 = W/\sqrt{2}$.
We have shown that the scalar potential following from $G$-invariance 
allows such a relation among the VEVs of the scalar gauge singlets,
apart from corrections suppressed by $(v/W)^2$,
where $v \simeq$ 246 GeV.
If $W$ is of the order of the seesaw scale,
then these corrections are completely negligible.
It is,
consequently,
also natural that the physical scalars associated with $\chi_{1,2}$
have masses of similar order of magnitude.
Due to the form of the neutrino mass matrix in Eq.~(\ref{Mnu}),
the solar mixing angle is free,
but large and non-maximal if we do no employ finetuning.

Since there are two scales in the scalar potential of our model,
the electroweak scale and $m_R$,
the scale of the right-handed-neutrino Majorana masses,
the model suffers from the naturalness problem associated with
radiative corrections to the masses of the light scalars.
Those masses receive radiative corrections of order $m_R^2$.
A high level of finetuning is required in order to eliminate
those large radiative corrections through renormalization.
A possible solution to this problem,
without destroying the nice features of the model,
might be supersymmetrization.

Our model requires a neutrino mass spectrum with $m_1 < m_2 < m_3$.
In addition,
there is a lower bound on $m_1$ of about $3 \times 10^{-3}\, \mathrm{eV}$,
if we use the best-fit values for the solar mixing angle
and for the mass-squared differences
$\Delta m^2_\mathrm{atm}$ and $\Delta m^2_\odot$.
Neutrinoless $\beta\beta$ decay is not necessarily
within the sensitivity range of future experiments,
but it may well be,
cf.\ Eq.~(\ref{r2}).
Among lepton-flavor-violating processes,
the same applies to the decay $\mu^\pm \to e^\pm e^+ e^-$,
while $\mu^\pm \to e^\pm \gamma$ is strongly suppressed;
lepton-flavor violation is just the same as in the model
with soft lepton-flavor breaking of Refs.~\cite{GL01,GL02}.

\vspace*{10mm}

\noindent {\bf Acknowledgement} The work of L.L.\ was supported by
\textit{Funda\c c\~ao para a Ci\^encia e a Tecnologia} (Portugal)
under the contract CFIF-Plurianual.

\end{document}